\documentclass[12pt]{article}
\usepackage[dvips]{graphicx}
\usepackage{amssymb}
\usepackage{amsmath}

\def\beq{\begin{equation}}\def\eeq{\end{equation}}
\def\bea{\begin{eqnarray}}\def\eea{\end{eqnarray}}

\begin{document}

\title{The use of test functions to help define quadratic Lagrangian on a causal set}
\author{Roman Sverdlov,
\\Department of Mathematics, University of New Mexico} 

\date{July 12, 2018}
\maketitle

\begin{abstract}
In some other papers, the Lagrangians in the causal sets included coefficients that were to be computed by integrating over Alexandrov set. In those other papers, this integral was explicitly evaluated, which resulted in rather sophisticated expressions. On the other hand, in this paper, instead of evaluating this integral, it was left in an integral form where the actual fields were replaced with test functions (thus avoiding nonlinearities). The test functions get absorbed into equations in a very natural way, so the resulting formulas look elegant.  \end{abstract}

\subsection*{Introduction}

A \emph{causal set} is a discrete model of spacetime (for reviews, see \cite{Review1} and \cite{Review2}). Namely, it is a partially ordered set, where partial ordering, $\prec$, represents lightcone causal relation: $p \prec q$ if and only if $p$ is in the past lightcone of $q$, or, equivalently, $q$ is in the future lightcone of $p$. This was  motivated by observations in \cite{Hawking} and \cite{Malament} that the metric of the Riemannian manifold, up to conformal scaling, can be identified by lightcone causal relations. An \emph{Alexadrov set} $\alpha (p,q)$ is defined to be 
\beq p \prec q \Longrightarrow \alpha (p,q) = \{r \vert p \prec r \prec q \} \eeq 
A \emph{distance} between $p$ and $q$ is defined as (see \cite{TimelikeDistance})
\beq p \prec q \Longrightarrow \tau (p, q) = \xi_0 \max \{n \vert \exists r_1, \cdots, r_{n-1} (p \prec r_1 \prec \cdots \prec r_{n-1} \prec q \} \label{Distance} \eeq 
where $\xi_0$ is some atomic length scale, such as Planck scale. 

In a series of papers (\cite{LagrangianGenerators},\cite{GravityAndMatter}, \cite{Gauge}) it was suggested that Alexandrov sets can be used in defining Lagrangian density. We are assuming that the relevant field is smooth and Alexandrov set is small. Consequently, the field is linear in the reference frame of the Alexandrov set. If we align $t$-axis with the axis of the Alexandrov set, we obtain 
\beq \partial_0 \phi \approx \frac{\phi (q) - \phi (p)}{\tau (p,q)} \eeq
\beq \frac{1}{\tau^{d+2} (p,q)} \int_{\alpha (p,q)} d^d r (\phi (r) - \phi (p))^2 = A^{scal}_1 (d) (\partial_0 \phi)^2 + A^{scal}_2 (d) \sum_k (\partial_k \phi)^2 \eeq
Consequently, one can find $C(d)$ for which 
\beq \partial^{\mu} \phi \partial_{\mu} \phi = N_{scal} \bigg[ \bigg(\frac{\phi (q) - \phi (p)}{\tau (p,q)} \bigg)^2 - \frac{C_{scal} (d)}{\tau^{d+2} (p,q)} \int_{\alpha (p,q)} d^d r (\phi (r) - \phi (p))^2 \bigg] \label{ScalarIntro} \eeq
for appropriately chosen $C(d)$. Furthermore, if one defines gauge field in terms of scalar two-point function, 
\beq a (p,q) = \int_{\gamma (p,q)} A_{\mu} dx^{\mu} \eeq
and again use the $pq$-segment as a $t$-axis, one can show that 
\beq \int_{\alpha (p,q)} d^d r (a(p,r) + a(r,q) + a(q,p))^2 \approx A_3 (d) \tau^{d+2} (p,q) \vert \vec{E} \vert^2 \eeq
\beq \int_{\alpha (p,q)} \bigg( d^d r \int_{\alpha (r,q)} \Big( d^d s (a(p,r) + a(r,s) + a(s,q)+ a(q,p))^2 \Big) \bigg) \approx \nonumber \eeq
\beq \approx \tau^{2d+2} (p,q) ( A_4 (d)  \vert \vec{E} \vert^2 + A_5 (d)  \vert \vec{B} \vert^2) \eeq
and, consequently, 
\beq F^{\mu \nu} F_{\mu \nu} \approx N_{EM} \bigg[ \frac{1}{\tau^{d+2} (p,q)} \int_{\alpha (p,q)} d^d r (a(p,r) + a(r,q) + a(q,p))^2 - \nonumber \eeq
\beq - \frac{C_{EM} (d)}{\tau^{2d+2}(p,q)} \int_{\alpha (p,q)} \bigg( d^d r \int_{\alpha (r,q)} \Big( d^d s (a(p,r) + a(r,s) + a(s,q)+ a(q,p))^2 \Big) \bigg) \bigg] \label{EMIntro} \eeq
By inspection of Eq \ref{ScalarIntro} and Eq \ref{EMIntro}, one can deduce that, for a general field $\cal F$, the Lagrangian takes the form 
\beq {\cal L} ({\cal F}; p, q) = N ({\cal K}_1 ({\cal F}; p,q) - C(d) {\cal K}_2 ({\cal F}; p,q)) \label{Generator} \eeq
However, in light of the fact that $\alpha (p,q)$ consists of cones as opposed to cubes, the expressions for $C(d)$ can become quite complicated. If the coefficient is selected to be anything other than $C(d)$ then the contributions coming from space and time coordinates won't combine in the right way, and the Lagrangian density would pick up frame-dependence, where the role of ``preferred frame" would be played by the geodesic connecting $p$ and $q$. This might seem a bit odd: why wouldn't causal set, which is defined in Lorentz invariant way, naturally produce invariant Lagrangian? The answer presented in \cite{Edges} is that the Lagrangian with ``wrong" coefficients would still be Lorentz invariant, just in a different sense. In particular, we can identify the direction of the line connecting $p$ and $q$ with a tangent vector, $v$, to a manifold. Thus, ${\cal L} ({\cal F}; p, q)$ is not a function over a manifold but, instead, it is a function over a tangent bundle. This being the case, its $v$-dependence in relation to Lorentz invariance would be no more disturbing than $x$-dependence in relation to translational invariance. The reason we are still trying to get rid of $v$-dependence by properly adjusting $C(d)$ has nothing to do with trying to save the invariance but, rather, it has to do with the fact that such $v$-dependence was not detected in the lab (in contrast to $x$-dependence that was) so we are trying to match experimental evidence. 

Be it as it may, it is still rather odd that the complicated expression for $C(d)$ had to be computed. For one thing, the theory is meant to be applicable for \emph{all} partially ordered sets, and most of them aren't manifold-like. So what $d$ would we pick if the partially ordered set wasn't manifoldlike? The answer to this question is that $d$ is not a dimension, its just an integer, which means that we can use that same $d$. It just \emph{happened} that the dimension of the manifoldlike causal set \emph{happened} to coincide with $d$, and thats why we ended up canceling $v$-dependence, but it didn't have to be the case. To emphasize this point, if we were living in a manifold of dimension $d' \neq d$, we would still be using $C (d)$ \emph{as opposed to} $C(d')$. After all, $C(d)$ can be used for \emph{all} causal sets (including non-manifoldlike ones), so why should a manifoldlike causal set of dimension $d' \neq d$ be an exception? As a result, in a causal set of dimension $d' \neq d$ we would obtain observable $v$-dependence of Lagrangian density, and we were simply lucky to be living in the universe of dimension $d$, which is the unique dimension where this doesn't happen. And we were \emph{even more lucky} that the formula for $C(d)$ that nature picked exactly coincided with the complicated expression given in \cite{Edges}, which is what needed to happen for such dimension to exist on the first place. 

As one can see, this argument is rather odd. In order to avoid this, it was suggested in \cite{LagrangianGenerators} that $C$ is not a constant but, instead, it keeps changing based on the local behavior of $\cal F$: nature simply checks by trial and error which $C$ will minimize the variations of Lagrangian density when between different sample Alexandrov sets flowing around, and picks that minimal value. In case of dimension $d$, it would approximate $C(d)$, and then if it changes to $d'$ it would, with some inertia, change to $C(d')$ (although in neither case it would be exactly equal to those quantities due to discreteness as well as the fact that $\cal F$ is only approximately linear on a small region, not exactly). This approach, however, has a serious weakness. An eventual goal of proposing an action should be the evaluation of a path integral. Even though we might not be able to do it due to lack of knowledge of the exact structure of a causal set, we would like to be able to compute it \emph{in a hypothetical scenario} if we knew the exact structure. For that reason, we would like the Lagrangian to be quadratic function of $\cal F$, which it is, with $C (d)$ being one of the coefficients. But, if $C$ were to pick up an $\cal F$-dependence, this would no longer be a quadratic function and, therefore, we would no longer be able to compute path integral without resorting to numeric simulations.

In order to avoid this difficulty, in this paper we propose a different approach. We introduce a \emph{test function} $f$, and we will use $f$, rather than $\cal F$, in order to find out what the coefficient $C$ should be. The test function $f$ is $\cal F$-independent, and, therefore, the resulting Lagrangian is quadratic in $\cal F$. If, however, we were to use only one such function $f$, it would, in spirit, be similar to a preferred frame, which contradicts the spirit of causal set theory. For that reason, we will average over several different trajectories of $f$. At the same time, we don't want to take a full blown path integral over $f$; instead, we will define a correspondence $(P,Q) \mapsto f_{PQ}$, where $P$ and $Q$ represents pair of points. We then average over all choices of $P$ and $Q$. It is important to stress that, for appropriately chosen $P$ and $Q$, that one $f_{PQ}$ will be sufficient for an approximation. The reason we do the averaging is to avoid any pair of points being preferred. 

\subsection*{The treatment of generalized fields}

As we just said, one choice of test function is sufficient for approximation. So lets start by assuming just one test function $f$, and then go back to the averaging shortly thereafter. Our goal is to find the value of $C$ for which $v$-dependence is negligibly small. We will take the following approach: we will select two Alexandrov sets $\alpha (p_1, q_1)$ and $\alpha (p_2, q_2)$ that are Lorentz boosts of each other (that is, $\tau (p_1, q_1) = \tau (p_2, q_2)$ and they have common midpoint) and find the value of $C$ for which the estimates of the Lagrangian density corresponding to those two Alexandrov sets approximate each other. In the case of $d$ dimensions, such $C$ would approximate $C(d)$.

In order for this argument to go through, $f$ has to be approximately linear both throughout $\alpha (p_1,q_1)$ and $\alpha (p_2,q_2)$. On the first glance, approximate linearity is merely a consequence of smoothness of $f$ and the smallness of the size of the Alexandrov sets. However, any given $f$ which we regard as smooth would no longer be smooth in a reference frame moving sufficiently close to the speed of light (after all, Lorentz boost will make $\partial^2_0 f$ large, and, in light of discreteness, the situation where $\partial^2_0 f > 1/ \xi$ corresponds to the situation where $f$ is no longer smooth). Since two randomly selected reference frames move arbitrary close to the speed of light relative to each other, $f$ is smooth only in a small minority of frames (even though said ``small minority" includes the range $(-0.99c, 0.99c)$). Now, if the velocity of either $\alpha (p_1, q_1)$ or $\alpha (p_2, q_2)$ were very close to the speed of light with respect to those frames, that(those) Alexandrov set(s) would look very long and very thin, which again would invalidate the assumption of linearity. Thus, the velocities of $\alpha (p_1, q_1)$ and $\alpha (p_2, q_2)$ with respect to the frames where $f$ is smooth, as well as with respect to each other, should be sufficiently smaller than $c$. The sets of all possible $(p_1, q_1, p_2, q_2, f)$ that meet this condition will be denoted by $S$. Thus,  
\beq (p_1,q_1,p_2,q_2, f) \in S \Longrightarrow \nonumber \eeq
\beq \Longrightarrow {\cal K}_1 (f; p_1, q_1) - C(d) {\cal K}_2 (f; p_1, q_1) \approx {\cal K}_1 (f; p_2, q_2) - C(d) {\cal K}_2 (f; p_2, q_2) \eeq
We can rearrange the terms in the above equation to obtain
\beq (p_1,q_1,p_2,q_2,f) \in S \Longrightarrow \nonumber \eeq
\beq \Longrightarrow {\cal K}_1 (f; p_1, q_1) -  {\cal K}_1 (f; p_2, q_2) \approx C(d) ({\cal K}_2 (f; p_1, q_1)- {\cal K}_2 (f; p_2, q_2)) \eeq
This can be further rewritten as 
\beq   (p_1,q_1,p_2,q_2,f) \in S \Longrightarrow C(d) \approx \frac{ {\cal K}_1 (f; p_1, q_1) -  {\cal K}_1 (f; p_2, q_2)}{{\cal K}_2 (f; p_1, q_1)- {\cal K}_2 (f; p_2, q_2)} \eeq
and substituting this into Eq \ref{Generator} gives
\beq (p_1,q_1,p_2,q_2, f) \in S \Longrightarrow \nonumber \eeq
\beq \Longrightarrow {\cal L} ({\cal F}; p_0, q_0) \approx N \bigg({\cal K}_1 ({\cal F}; p_0,q_0) - \frac{ {\cal K}_1 (f; p_1, q_1) -  {\cal K}_1 (f; p_2, q_2)}{{\cal K}_2 (f; p_1, q_1)- {\cal K}_2 (f; p_2, q_2)}  {\cal K}_2 ({\cal F}; p,q) \bigg)  \label{GlobalVersion} \eeq
where $p$ and $q$ in Eq \ref{Generator} were replaced with $p_0$ and $q_0$. Note that $(p_1, q_1, p_2, q_2, f) \in S$ makes no mention of $p_0$ and $q_0$, which implies that $p_0$ and $q_0$ can be arbitrarily far away from the other points: after all, $p_1$, $q_1$, $p_2$ and $q_2$ play no physical role other than telling us what $C(d)$ is, which is constant throughout the manifold.  Of course, the assumption we made is that $\cal F$ is well behaved in the reference frame of $\gamma (p_0,q_0)$, which is why it is approximately linear throughout $\alpha (p_0,q_0)$. But, as we said earlier, $\cal F$ and $f$ are independent of each other and, therefore, none of this implies any relation between the location of $\alpha (p_0,q_0)$ and the location of the other Alexandrov sets.

Now it is time to go back to averaging over multiple choices of $f$. We introduce a weight function $W(p_0,q_0,p_1,q_1,p_2,q_2,f)$ in such a way that 
\beq  \forall (p_1,q_1,p_2,q_2,f) \not\in T  (W(p_1,q_1,p_2,q_2,f) = 0 ) \eeq
which assures us that the only contributions to weighted average come from $S$. In order to be able to take weighted sum, let us pretend for a minute that causal set is bounded -- and then we will abandon that assumption shortly thereafter. So, the Lagrangian estimate is given by 
\beq  {\cal L} ({\cal F}; p_0, q_0) \approx \frac{N}{\sum_{p'_1,q'_1,p'_2,q'_2,g} W(p'_1,q'_1,p'_2,q'_2,g)} \times  \eeq
\beq \times \sum_{p_1,q_1,p_2,q_2,f} W(p_1,q_1,p_2,q_2,f) \bigg({\cal K}_1 ({\cal F}; p_0,q_0) - \frac{ {\cal K}_1 (f; p_1, q_1) -  {\cal K}_1 ({\cal F}; p_2, q_2)}{{\cal K}_2 (f; p_1, q_1)- {\cal K}_2 (f; p_2, q_2)}  {\cal K}_2 (f; p,q) \bigg) \nonumber \eeq
Lets define a different normalization constant, $M$, as well as different weight function, $w$, as follows:
\beq M =  \frac{N}{\sum_{p'_1,q'_1,p'_2,q'_2,g} W(p'_1,q'_1,p'_2,q'_2,g)} \eeq
\beq w (p_1, q_1, p_2, q_2,f) = \frac{W( p_1, q_1, p_2, q_2, f)}{{\cal K}_2 (f; p_1, q_1)- {\cal K}_2 (f; p_2, q_2)} \eeq
Then the Lagrangian becomes 
\beq {\cal L} ({\cal F}; p_0, q_0) \approx M \sum_{p_1,q_1,p_2,q_2,f} \bigg( w(p_1,q_1,p_2,q_2,f) \times \nonumber \eeq
\beq \times \Big({\cal K}_1 ({\cal F}; p_0, q_0) \big({\cal K}_2 (f; p_1, q_1) - {\cal K}_2 (f; p_2, q_2) \big)  - \nonumber \eeq
\beq - {\cal K}_2 ({\cal F}; p_0, q_0) \big({\cal K}_1 (f; p_1, q_1) - {\cal K}_1 (f; p_2, q_2) \big) \Big) \bigg) \eeq
We would now like to abandon the assumption that causal set is bounded. In order to do that, we replace $w(p_1, q_1, p_2, q_2, f)$ with $w(p_0,q_0, p_1, q_1, p_2, q_2, f)$. In order for it to be non-zero, the previous conditions pertaining $p_1$, $q_1$, $p_2$, $q_2$ and $f$ have to meet, and, in addition to that, the distances from $p_0$ and $q_0$ to the rest of the points have to be smaller than certain agreed-upon constants. As mentioned earlier, in the manifold case, said upper bound has no physical meaning, and its only purpose is the technicality of allowing us to find the average when the space is unbounded. However, the situation in which this becomes more important is when we replace the assumption that the space is $d$-dimensional throughout with a weaker assumption that its topology is ``slowly changing" -- and, therefore, it is $d$-dimensional on some neighborhood of $p_0$ and $q_0$. In this case, the rest of the points better fall within that neighborhood. In any case, the new, generalized, equation becomes 
\beq {\cal L} ({\cal F}; p_0, q_0) \approx M \sum_{p_1,q_1,p_2,q_2,f} \bigg( w(p_0,q_0,p_1,q_1,p_2,q_2,f) \times \nonumber \eeq
\beq \times \Big({\cal K}_1 ({\cal F}; p_0, q_0) \big({\cal K}_2 (f; p_1, q_1) - {\cal K}_2 (f; p_2, q_2) \big)  - \nonumber \eeq
\beq - {\cal K}_2 ({\cal F}; p_0, q_0) \big({\cal K}_1 (f; p_1, q_1) - {\cal K}_1 (f; p_2, q_2) \big) \Big) \bigg) \eeq
As we mentioned earlier, we could have obtained the approximation by picking just one $f$, and the purpose of weighted average is somewhat aesthetic. This being the case, let us reduce the number of test functions we are considering by only looking at the ones generated by pairs of points, $(p_3, q_3) \mapsto f_{p_3q_3}$ (we introduce index $3$ in order to make sure that the pair of points we are using to generate $f$ is distinct from any other pair of points we were using). On the one hand, this simplifies the model, since the number of pairs of points is much smaller than the number of trajectories of $f$ we would consider well behaved; on the other hand, the number of pairs of points is still large enough for us to avoid the appearance of preferred reference frame. Thus, we replace $w(p_0,q_0,p_1,q_1,p_2,q_2,f)$ with $w(p_0,q_0,p_1,q_1,p_2,q_2,p_3,q_3)$ (that is, we replaced $f$ with $p_3,q_3$ in the list of the variables of $w$) and then replace ${\cal K}_i (f;p_j,q_j)$ with ${\cal K}_i (f_{p_3,q_3}; p_j, q_j)$ (that is, replace $f$ with $f_{p_3q_3}$ inside ${\cal K}_i (\cdots)$). This gives us
\beq {\cal L} ({\cal F}; p_0, q_0) \approx M \sum_{p_1,q_1,p_2,q_2,p_3,q_3} \bigg( w(p_0,q_0,p_1,q_1,p_2,q_2,p_3,q_3) \times \nonumber \eeq
\beq \times \Big({\cal K}_1 ({\cal F}; p_0, q_0) \big({\cal K}_2 (f_{p_3q_3}; p_1, q_1) - {\cal K}_2 (f_{p_3q_3}; p_2, q_2) \big)  - \nonumber \eeq
\beq - {\cal K}_2 ({\cal F}; p_0, q_0) \big({\cal K}_1 (f_{p_3q_3}; p_1, q_1) - {\cal K}_1 (f_{p_3q_3}; p_2, q_2) \big) \Big) \bigg) \label{LagrangianSimpleGeneral} \eeq 
To streamline the notation, lets define a \emph{truth function}, $T$, that would give value of $1$ to all true statements and value of $0$ to all the false statements: 
\beq T (true) = 1 \; , \; T (false) = 0 \eeq
The weight function can be defined as 
\beq w(p_0,q_0,p_1,q_1,p_2,q_2,p_3,q_3) = T(\forall i, j \in \{0, \cdots, 3 \} (p_i \prec q_j)) \times \nonumber \eeq
\beq \times T(\forall i ((p_i \prec p_3) \vee (p_3 \prec p_i)) T(\forall i ((q_i \prec q_3) \vee (q_3 \prec q_i)) \times  \nonumber \eeq
\beq \times T \bigg(\exists r \bigg(\tau (p_1,r) = \tau (p_2, r) = \tau (r, q_1) = \tau (r, q_2) = \frac{\tau (p_1, q_1)}{2} = \frac{\tau(p_2,q_2)}{2} \bigg) \bigg) \times \nonumber \eeq
\beq \times \prod_{i=0}^3 \prod_{j=0}^3 T(l_{ij} \leq \tau (p_i,q_j) \leq L_{ij}) \label{WeightFromDistance} \eeq
where $l_{ij}$ and $L_{ij}$ are some agreed upon constants. The purpose of the right hand side on the first line is to make sure that all the distances that are mentioned on the other two lines are, indeed, well defined: after all, the distance given in Eq \ref{Distance} is defined only for causally related pairs of points. The purpose of the second line is similar. Whereas for $i,j \in \{0,1,2 \}$ we are content with just knowing distances between $p_i$ and $q_j$, in case of $3$ we want to know the distance from $p_3$ to \emph{both} $p_i$ and $q_i$ in order to utilize it for the definition of $f_{p_3q_3}$ (see Eq \ref{fEM} and \ref{fScal}) since the relation between $p$ and $q$ was already taken care in the first line, thats why the second line only deals with $p$ and $p$ as well as $q$ and $q$. In cases of both first and second line, If we wanted, we could have used \cite{Distance} to define distances between spacelike separated points, and then we would no longer need those causal relations. But the definition presented in \cite{Distance} involves infinite averaging over the Lorentz boosts on the hyperplane orthogonal to the spacelike line in question. So, in order to avoid unnecessary controversy, we prefer to only use timelike distances. The purpose of the third line is to say that the segments have a common midpoint, which is what implies that they are Lorentz boosts of each other; also, the last equal sign on the third line implies that the two Alexandrov sets have the same size, which is important since what we are computing is really a product of Lagrangian density with the volume of Alexandrov set, so they better have the same size in order for us to anticipate that those products are equal. The purpose of the fourth line is to prevent the situations where the Lorentz stretching makes one or both Alexandrov sets very long and very thin in reference frame of the third one, as well as to make sure that $\alpha (p_0, q_0)$ isn't displaced too far away from the other Alexandrov sets, and thus avoid infinite sum during averaging (it also avoids the situation of $p_3$ or $q_3$ being displaced too far, for that same reason). 

In case of multiple fields, such as ${\cal F}_1$ and ${\cal F}_2$, one might ask the following question: should we add ${\cal F}_1$ and ${\cal F}_2$ and then compute a single Lagrangian for the sum, or should we compute separate Lagrangians for ${\cal F}_1$ and ${\cal F}_2$ first, and then add them? A quick inspection of the papers that explicitly compute $C(d)$ tells us that in case of two fields we would have two separate coefficients, $C_1 (d)$ and $C_2 (d)$ that won't be equal to each other (in fact, their expressions are quite different). This being the case, one has to \emph{first} compute two separate Lagrangians and \emph{then} add them. Each of those Lagrangians might likely have separate weight functions as well, which we denote by $w_1$ and $w_2$. Thus, 
\beq {\cal L} ({\cal F}; p_0, q_0) \approx \nu_1 \sum_{p_1,q_1,p_2,q_2,p_3,q_3} \bigg( w_1(p_0,q_0,p_1,q_1,p_2,q_2,p_3,q_3) \times \nonumber \eeq
\beq \times \Big({\cal K}_1 ({\cal F}; p_0, q_0) \big({\cal K}_2 (f_{p_3q_3}; p_1, q_1) - {\cal K}_2 (f_{p_3q_3}; p_2, q_2) \big)  - \nonumber \eeq
\beq - {\cal K}_2 ({\cal F}; p_0, q_0) \big({\cal K}_1 (f_{p_3q_3}; p_1, q_1) - {\cal K}_1 (f_{p_3q_3}; p_2, q_2) \big) \Big) \bigg) + \nonumber \eeq
\beq +  \nu_2 \sum_{p_1,q_1,p_2,q_2,p_3,q_3} \bigg( w_2(p_0,q_0,p_1,q_1,p_2,q_2,p_3,q_3) \times \nonumber \eeq
\beq \times \Big({\cal K}_1 ({\cal F}; p_0, q_0) \big({\cal K}_2 (f_{p_3q_3}; p_1, q_1) - {\cal K}_2 (f_{p_3q_3}; p_2, q_2) \big)  - \nonumber \eeq
\beq - {\cal K}_2 ({\cal F}; p_0, q_0) \big({\cal K}_1 (f_{p_3q_3}; p_1, q_1) - {\cal K}_1 (f_{p_3q_3}; p_2, q_2) \big) \Big) \bigg)\label{LagrangianSimpleGeneral2} \eeq 
where the difference between $w_1$ and $w_2$ amounts to replacing $l_i$ and $L_i$ with $l_{1i}$ and $L_{1i}$, in case of $w_1$, and replacing it with $l_{2i}$ and $L_{2i}$ in case of $w_2$:
\beq w_1(p_0,q_0,p_1,q_1,p_2,q_2,p_3,q_3) = T(\forall i, j \in \{0, \cdots, 3 \} (p_i \prec q_j)) \times \nonumber \eeq
\beq \times T(\forall i ((p_i \prec p_3) \vee (p_3 \prec p_i)) T(\forall i ((q_i \prec q_3) \vee (q_3 \prec q_i)) \times  \nonumber \eeq
\beq \times T \bigg(\exists r \bigg(\tau (p_1,r) = \tau (p_2, r) = \tau (r, q_1) = \tau (r, q_2) = \frac{\tau (p_1, q_1)}{2} = \frac{\tau(p_2,q_2)}{2} \bigg) \bigg) \times \nonumber \eeq
\beq \times \prod_{i=0}^3 \prod_{j=0}^3 T(l_{1ij} \leq \tau (p_i,q_j) \leq L_{1ij}) \label{WeightFromDistance1} \eeq
\beq w_2(p_0,q_0,p_1,q_1,p_2,q_2,p_3,q_3) = T(\forall i, j \in \{0, \cdots, 3 \} (p_i \prec q_j)) \times \nonumber \eeq
\beq \times T(\forall i ((p_i \prec p_3) \vee (p_3 \prec p_i)) T(\forall i ((q_i \prec q_3) \vee (q_3 \prec q_i)) \times  \nonumber \eeq
\beq \times T \bigg(\exists r \bigg(\tau (p_1,r) = \tau (p_2, r) = \tau (r, q_1) = \tau (r, q_2) = \frac{\tau (p_1, q_1)}{2} = \frac{\tau(p_2,q_2)}{2} \bigg) \bigg) \times \nonumber \eeq
\beq \times \prod_{i=0}^3 \prod_{j=0}^3 T(l_{2ij} \leq \tau (p_i,q_j) \leq L_{2ij}) \label{WeightFromDistance2} \eeq
Of course, there are more than two fields, in which case the sum of two terms would be replaced with the sum of $a$ terms, which we don't want to write in order not to mislead the reader, since the number of terms is still small (for example, for the standard model with gravity it would only be $4$ terms).

Notice that, even though we were using the manifold geometry throughout our reasoning, the last two equations, in conjunction with Eq \ref{Distance}, formally make sense in a general partially ordered set. Therefore, we claim that those two equations are fundamental, while the differential equations we know in continuum are emergent.  It is still true that  ${\cal L}_{continuum} \approx {\cal L}_{discrete}$ is an approximation. But it is compatible with two situations: ${\cal L}_{discrete} \approx {\cal L} = {\cal L}_{continuum}$ and ${\cal L}_{discrete} = {\cal L} \approx {\cal L}_{continuum}$. Thus, we make sure to believe the latter rather than the former and, accordingly, change approximation sign in Eq \ref{WeightFromDistance} into an exact equal sign: 
\beq {\cal L} ({\cal F}; p_0, q_0) = \nu_1 \sum_{p_1,q_1,p_2,q_2,p_3,q_3} \bigg( w_1(p_0,q_0,p_1,q_1,p_2,q_2,p_3,q_3) \times \nonumber \eeq
\beq \times \Big({\cal K}_1 ({\cal F}; p_0, q_0) \big({\cal K}_2 (f_{p_3q_3}; p_1, q_1) - {\cal K}_2 (f_{p_3q_3}; p_2, q_2) \big)  - \nonumber \eeq
\beq - {\cal K}_2 ({\cal F}; p_0, q_0) \big({\cal K}_1 (f_{p_3q_3}; p_1, q_1) - {\cal K}_1 (f_{p_3q_3}; p_2, q_2) \big) \Big) \bigg) + \nonumber \eeq
\beq +  \nu_2 \sum_{p_1,q_1,p_2,q_2,p_3,q_3} \bigg( w_2(p_0,q_0,p_1,q_1,p_2,q_2,p_3,q_3) \times \nonumber \eeq
\beq \times \Big({\cal K}_1 ({\cal F}; p_0, q_0) \big({\cal K}_2 (f_{p_3q_3}; p_1, q_1) - {\cal K}_2 (f_{p_3q_3}; p_2, q_2) \big)  - \nonumber \eeq
\beq - {\cal K}_2 ({\cal F}; p_0, q_0) \big({\cal K}_1 (f_{p_3q_3}; p_1, q_1) - {\cal K}_1 (f_{p_3q_3}; p_2, q_2) \big) \Big) \bigg)\label{LagrangianSimpleGeneral3} \eeq 

\subsection*{A specific example} 

 Let us now apply what we have learned to a specific example. We will pattern this example after the model presented in \cite{Edges}, but the reader should keep in mind that similar procedure can be used to pattern an example after any other proposals (such as \cite{GravityAndMatter}, \cite{Gauge}, and so forth). Our example will be a toy model with electromagnetic field and a charged scalar field. 

 First of all, we would like to define $b_{p_3q_3}$. Since we don't have any clues on what it is supposed to be, we will just pull something out of the air. The only condition, though, is that it should produce non-zero $F_{\mu \nu}$. If we were to have just one point $p_3$, it would have been impossible without violating the symmetry around $p_3$. That is why we introduce two points $p_3$ and $q_3$ (in case of scalar field discussed in the next section two points are no longer necessary; we just kept two points in the general presentation to accommodate electromagnetic case). While $A_{\mu}=\partial_{\mu} \Lambda$ corresponds to $F_{\mu \nu} = 0$, replacing it with $A_{\mu}= \chi \partial_{\mu} \Lambda$ wouldn't: 
\beq A_{\mu} = \chi \partial_{\mu} \Lambda \Longrightarrow F_{\mu \nu} = \partial_{\mu} \chi \partial_{\nu} \Lambda - \partial_{\nu} \chi \partial_{\mu} \Lambda \label{GradientsNotParallel}\eeq
which is non-zero \emph{as long as $\partial_{\mu} \chi$ and $\partial_{\mu} \Lambda$ aren't colinear}. In order to insure the latter, we have to assign the sources to $\chi$ and $\Lambda$ to $p_3$ and $q_3$, respectively, and then separate them by a big distance from each other, which amounts to saying 
\beq l^{EM}_{33} \gg \max \{L_{ij} \vert \{i,j \} \subset \{1,2 \} \} \label{l33} \eeq
We then define 
\beq \chi (r) = \tau (p_3, r) \; , \; \Lambda (r) = \tau (q_3, r) \eeq
and, if we denote electromagnetic version of $f_{pq}$ by $b_{pq}$, then Eq \ref{GradientsNotParallel} can be rewritten as 
\beq b_{pq}(r,s) = \frac{1}{2} (\tau (p,r) + \tau (p,s))(\tau (q,s)- \tau (q,r)) \label{fEM} \eeq
In order for $a$ to be well behaved, we need to avoid being at the vicinity of $p_3$ and $q_3$. Thus, we assume 
\beq \forall i \in \{0,1,2,3 \} (\min \{l_{i3}, l_{3i}\} \gg \max \{L_{ij} \vert \{i,j \} \subset \{1,2 \} \} ) \label{li3} \eeq
and, since we deliberately included $i=3$, this includes Eq \ref{l33} as a special case. 

While gauge field is modeled in terms of two-point function, the scalar field, in most approaches, is modeled as one point function (see, for example, \cite{Dambertian1}, \cite{Dambertian2}, \cite{GravityAndMatter}, \cite{Bosonic} and \cite{Nonlinear} for some alternative approaches). However, a common problem with those approaches is nonlocality of Lorentzian neighborhood of any given point (see \cite{Nonlocality} and \cite{NonlocalityTheorem}). In light of the fact that Lorentzian neighborhood is a vicinity of light cone, which has infinite volume, a point would have infinitely many direct neighbors. In most other discrete theories this situation is avoided because the specific discrete structure selects a preferred frame. For example, in case of a cubic lattice, the edges of the lattice would determine preferred direction. But, since causal set is modeled in terms of Poisson distribution, such is not the case. One way out of this predicament was suggested in \cite{Edges}: namely, if \emph{all} fields, including scalar field, reside on edges rather than points, then the preferred direction would be the direction of the edge. Thus, in \cite{Edges}, the scalar field was defined as a two-point function. In case of the embedding into continuum, this two point function is  
\beq \phi (p,q) = \frac{1}{l (\gamma (p,q))} \int_{\gamma (p,q)} \phi (r) d r \eeq
where $l (\gamma (p,q))$ is the length of the continuous curve $\gamma (p,q)$, not to be confused with discretized $\tau (p,q)$.  As with electromagnetic field, we then abandon the continuum picture on the right hand side, and just retain the left hand side as the basic entity. Unlike electromagnetic case, the sample function can be generated by just one point, $p_3$: 
\beq \eta_{p_3} (r,s) = \tau (p_3, s) - \tau (p_3, r) \label{fScal} \eeq
where $\eta_{pq}$ is a scalar version of $f_{pq}$. For formalities sake, we can simply make $q_3$ arbitrary and just say 
\beq \eta_{p_3q_3} (r,s) =\eta_{p_3} (r,s) \eeq
Now, we have two separate pairs of terms in the Lagrangians, one pair for electromagnetic field and the other one for scalar field:  
\beq {\cal K}^{EM}_1 (a; p,q) = \sum_{r \in \alpha (p,q)} \; \; \; \sum_{s \in \alpha (r,q)} (a(p,r) + a(r,s)+a(s,q)+a(q,p))^2 \label{KEM1} \eeq
\beq {\cal K}^{EM}_2 (a;p,q) = \sum_{\{r,s \} \subset \alpha (p,q)} (a(p,r) + a (r,q) + a (q,s) + a (s,p))^2 \label{KEM2} \eeq
\beq {\cal K}^{scal}_1 (\phi, a; p,q) = \sum_{r \in \alpha (p,q)} \bigg\vert \bigg(1- \frac{i}{2} a(p,q) \bigg) \phi (r,q) - \bigg(1+ \frac{i}{2} a(p,q) \bigg) \phi (p,r) \bigg\vert^2 \label{KScal1} \eeq
\beq {\cal K}^{scal}_2 (\phi,a; p,q) = \nonumber \eeq
\beq = \sum_{r \in \alpha (p,q)} \bigg\vert \bigg(1- \frac{i}{4} (a(p,r)+a(q,r)) \bigg) \frac{\phi (p,r) + \phi (r,q)}{2} - \bigg(1+ \frac{i}{4} (a(p,r)+a(q,r)) \bigg) \phi (p,q) \bigg\vert^2 \label{KScal2} \eeq
where we were assuming that all the distances are small, which makes the values of $a$ small and, therefore $e^{ia} = 1+ ia + 0(a^2)$; we chose to assume $1+ia$ is exact while $e^{ia}$ is only an approximation, in order to avoid higher order couplings. How, the Lagrangian is given by 
\beq {\cal L} (a, \phi; p_0,q_0) = \nu_{EM}  \sum_{p_1,q_1,p_2,q_2,p_3,q_3} \Bigg\{w_{EM} (p_1,q_1,p_2,q_2,p_3,q_3) \times \nonumber \eeq
\beq \times \bigg[{\cal K}^{EM}_1 (a;p_0,q_0) \Big({\cal K}^{EM}_2 (b_{p_3,q_3};p_1,q_1)- {\cal K}^{EM}_2 (b_{p_3,q_3};p_2,q_2) \Big) -  \nonumber \eeq
\beq - {\cal K}^{EM}_1 (a;p_0,q_0) \Big({\cal K}^{EM}_2 (b_{p_3,q_3};p_1,q_1)- {\cal K}^{EM}_2 (b_{p_3,q_3};p_2,q_2) \Big) \bigg] \Bigg\} + \nonumber \eeq
\beq + \nu_{scal}  \sum_{p_1,q_1,p_2,q_2,p_3,q_3} \Bigg\{ w_{scal} (p_1,q_1,p_2,q_2,p_3,q_3) \times \nonumber \eeq
\beq \times \bigg[{\cal K}^{scal}_1 (a, \phi;p_0,q_0) \Big({\cal K}^{scal}_2 (a, \eta_{p_3,q_3};p_1,q_1)- {\cal K}^{scal}_2 (a, \eta_{p_3,q_3};p_2,q_2) \Big) -  \nonumber \eeq
\beq - {\cal K}^{scal}_1 (a, \phi;p_0,q_0) \Big({\cal K}^{scal}_2 (a, \eta_{p_3,q_3};p_1,q_1)- {\cal K}^{scal}_2 (a, \eta_{p_3,q_3};p_2,q_2) \Big) \bigg]  \Bigg\} \eeq
where ${\cal K}^{EM}_1$, ${\cal K}^{EM}_2$, ${\cal K}^{scal}_1$, and ${\cal K}^{scal}_2$ are given by Eq \ref{KEM1}, \ref{KEM2}, \ref{KScal1} and \ref{KScal2}, respectively, $\nu_{scal}$ and $\nu_{EM}$ are constants, and $w_{EM}$ and $w_{scal}$ are given by taking Eq \ref{WeightFromDistance} and then replacing $l_i$ and $L_i$ with $l_i^{scal}$, $l_i^{EM}$, $L_i^{scal}$ and $L_i^{EM}$: 
\beq w_{EM} (p_0,q_0,p_1,q_1,p_2,q_2,p_3,q_3) = T(\forall i, j \in \{0, \cdots, 3 \} (p_i \prec q_j)) \times \nonumber \eeq
\beq \times T(\forall i ((p_i \prec p_3) \vee (p_3 \prec p_i)) T(\forall i ((q_i \prec q_3) \vee (q_3 \prec q_i)) \times  \nonumber \eeq
\beq \times T \bigg(\exists r \bigg(\tau (p_1,r) = \tau (p_2, r) = \tau (r, q_1) = \tau (r, q_2) = \frac{\tau (p_1, q_1)}{2} = \frac{\tau(p_2,q_2)}{2} \bigg) \bigg) \times \nonumber \eeq
\beq \times \prod_{i=0}^3 \prod_{j=0}^3 T(l^{EM}_{ij} \leq \tau (p_i,q_j) \leq L^{EM}_{ij}) \label{WeightFromDistanceEM} \eeq

\beq w_{scal} (p_0,q_0,p_1,q_1,p_2,q_2,p_3,q_3) = T(\forall i, j \in \{0, \cdots, 3 \} (p_i \prec q_j)) \times \nonumber \eeq
\beq \times T(\forall i ((p_i \prec p_3) \vee (p_3 \prec p_i)) T(\forall i ((q_i \prec q_3) \vee (q_3 \prec q_i)) \times \nonumber \eeq
\beq \times T \bigg(\exists r \bigg(\tau (p_1,r) = \tau (p_2, r) = \tau (r, q_1) = \tau (r, q_2) = \frac{\tau (p_1, q_1)}{2} = \frac{\tau(p_2,q_2)}{2} \bigg) \bigg) \times \nonumber \eeq
\beq \times \prod_{i=0}^3 \prod_{j=0}^3 T(l^{scal}_{ij} \leq \tau (p_i,q_j) \leq L^{scal}_{ij}) \label{WeightFromDistanceScal} \eeq

\subsection*{Conclusion}

The main accomplishment of this paper is that we were able to provide a way of converting a generic model that includes a complicated coefficient $C(d)$ into a much more elegant equation that doesn't have any such coefficient. We have explicitly applied it to the model described in \cite{Edges}, but it can be equally easily applied to similar models described in \cite{GravityAndMatter}, \cite{Gauge} and so forth. The key idea of this paper is to use test fields in order to find out what $C(d)$ might be. On the first glance, the idea of test fields sound reminiscent to the idea of preferred frame, which would go counter to the spirit of causal set theory. But, after working it out, it turns out that the structure of the equations we obtain at the end looks quite natural and frame independent. 

It is important to note that there is a different class of approaches to QFT in causal sets, to which this paper is irrelevant. Those include propagator approach by Johnston (\cite{Johnston1} and \cite{Johnston2}), non-local D'ambertians where nonlocal effects are claimed to cancel (\cite{Dambertian1} and \cite{Dambertian2}) and also attempts to directly construct the wavelents instead of using Lagrangian (see \cite{PhaseSpacetime} and \cite{LongEdges}). However, those other approaches were mainly focused on the scalar field, and the attempts to describe electromagnetic field in \cite{PhaseSpacetime} and \cite{Bosonic} were both quite awkward. On the other hand, the papers that do utilize $C(d)$, in fact, describe electromagnetic field (see \cite{Gauge} and \cite{Edges}) but the coefficients $C(d)$ make those approaches look awkward. The purpose of this paper is to make the latter two approaches more natural by replacing $C(d)$ with a naturally looking construction. 

One thing to do for a future research is to apply the framework of this paper to the gravity part of \cite{GravityAndMatter} thereby obtaining equation for curvature devoid of $C(d)$. That is more easily said than done, since it would require one to define the notion of ``sources" of gravitational field -- which, contrary to the sources we were using, would no longer assume fixed metric. This would, of course, mean that we would still have to deal with well known conceptual problems of quantum gravity, just like we would have in \cite{GravityAndMatter}; we would simply avoid an extra issue of $C(d)$. Be it as it may, it might still be worth doing, and if/when such thing is done it would be nice to compare it to the approach described in \cite{Curvature}, which doesn't have $C(d)$ on the first place.

\end{document}